\documentclass[final,3p,times,twocolumn]{elsarticle}

\usepackage{graphicx}
\usepackage{rotating}
\usepackage{multirow}
\usepackage{amsmath}
\usepackage{amssymb}
\usepackage{lineno}
\usepackage{comment}
\usepackage{microtype}
\usepackage{hyperref}
\usepackage[table, dvipsnames]{xcolor}

\begin{document}
\begin{frontmatter}


\title{Characterization of the LUNA neutron detector array for the measurement of the $^{13}$C($\alpha$,$n$)$^{16}$O reaction}

\author[add1,add2,add3]{L. Csedreki\corref{cor1}}
\ead{laszlo.csedreki@lngs.infn.it}
\author[add1,add2,add3]{G.F. Ciani}
\author[add4]{ J. Balibrea-Correa}
\author[add4]{A. Best}
\author[add6]{M. Aliotta}
\author[add7]{F. Barile}
\author[add8]{D. Bemmerer}
\author[add1]{A.~Boeltzig}
\author[add9]{C. Broggini}
\author[add6]{C.G. Bruno}
\author[add9,add10]{A. Caciolli}
\author[add13]{F. Cavanna}
\author[add6]{T. Chillery}
\author[add13,add12]{P. Colombetti}  \author[add11,add14]{P. Corvisiero}
\author[add6]{T. Davinson}
\author[add9, add10]{R. Depalo} 
\author[add4]{A. Di Leva}
\author[add3]{Z. Elekes} 
\author[add11,add14]{F. Ferraro}
\author[add7,add15]{ E.\,M. Fiore} \author[add1]{A. Formicola}
\author[add3]{Zs. F\"ul\"op}
\author[add13,add12]{G. Gervino} 
\author[add16]{A. Guglielmetti} 
\author[add18]{C. Gustavino}
\author[add3]{Gy. Gy\"urky}
\author[add4]{G. Imbriani}
\author[add19]{Z. Janas}
\author[add1]{M. Junker} 
\author[add1]{I. Kochanek} 
\author[add20,add25]{M. Lugaro} 
\author[add9,add10]{P. Marigo} 
\author[add16]{E. Masha} 
\author[add19]{C. Mazzocchi}
\author[add9]{R. Menegazzo} \author[add7]{V. Mossa}
\author[add7,add21]{F.\,R.\ Pantaleo}  \author[add7]{V.Paticchio} 
\author[add7]{R. Perrino}
\author[add9,add10]{D.Piatti}
\author[add11,add14]{P. Prati}
\author[add7,add15]{L. Schiavulli}
\author[add8,add22]{K. St\"ockel}
\author[add1,add23]{O. Straniero} 
\author[add3]{T. Sz\"ucs} 
\author[add8,add22]{M.\,P.\ Tak\'acs}
\author[add24]{F.Terrasi} 
\author[add11]{S. Zavatarelli}

\cortext[cor1]{Corresponding author}

\address[add1]{Istituto Nazionale di Fisica Nucleare Laboratori Nazionali del Gran Sasso (LNGS), Via G. Acitelli 22, 67100 Assergi, Italy}
\address[add2]{Gran Sasso Science Institute, Viale F. Crispi 7, 67100, L'Aquila, Italy }
\address[add3]{Institute for Nuclear Research (Atomki), PO Box 51, 4001 Debrecen, Hungary}
\address[add4]{Istituto Nazionale di Fisica Nucleare, Sezione di Napoli $\&$ Universit\`a degli Studi di Napoli ``Federico II'', Dipartimento di Fisica ``E. Pancini'', Via Cintia 21, 80126 Napoli, Italy}
\address[add6]{SUPA, School of Physics and Astronomy, University of Edinburgh, Peter Guthrie Tait Road, EH9 3FD Edinburgh, United Kingdom}
\address[add7]{Istituto Nazionale di Fisica Nucleare, Sezione di Bari, Via E. Orabona 4, 70125 Bari, Italy}
\address[add8]{Helmholtz-Zentrum Dresden-Rossendorf, Bautzner Landstra\ss{}e 400, 01328 Dresden, Germany}
\address[add9]{Istituto Nazionale di Fisica Nucleare, Sezione di Padova, Via F. Marzolo 8, 35131 Padova, Italy}
\address[add10]{Universit\`a degli Studi di Padova, Via F. Marzolo 8, 35131 Padova, Italy}
\address[add13]{Istituto Nazionale di Fisica Nucleare, Sezione di Torino , Via P. Giuria 1, 10125 Torino, Italy}
\address[add12]{Universit\`a degli Studi di Torino, Via P. Giuria 1, 10125 Torino, Italy}
\address[add11]{Istituto Nazionale di Fisica Nucleare, Sezione di Genova, Via Dodecaneso 33, 16146 Genova, Italy}
\address[add14]{Universit\`a degli Studi di Genova, Via Dodecaneso 33, 16146 Genova, Italy}
\address[add15]{ Universit\`a degli Studi di Bari, Dipartimento Interateneo di Fisica, Via G. Amendola 173, 70126 Bari, Italy}
\address[add16]{Universit\`a degli Studi di Milano \& Istituto Nazionale di Fisica Nucleare, Sezione di Milano, Via G. Celoria 16, 20133 Milano, Italy} 
\address[add18]{Istituto Nazionale di Fisica Nucleare, Sezione di Roma, Piazzale A. Moro 2, 00185 Roma, Italy}
\address[add19]{Faculty of Physics, University of Warsaw, ul. Pasteura 5, 02-093 Warszawa, Poland}
\address[add20]{Konkoly Observatory, Research Centre for Astronomy and Earth Sciences, Konkoly Thege Miklós út 15-17, H-1121 Budapest, Hungary}
\address[add21]{Politecnico di Bari, Dipartimento Interateneo di Fisica, Via G. Amendola 173, 70126 Bari, Italy}
\address[add22]{ Technische Universit\"at Dresden, Institut f\"ur Kern- und Teilchenphysik, Zellescher Weg 19, 01069 Dresden, Germany}
\address[add23]{INAF Osservatorio Astronomico d'Abruzzo, Via Mentore Maggini, 64100 Teramo, Italy}
\address[add24]{Universit\`a degli Studi della Campania L. Vanvitelli, Dipartimento di Matematica e Fisica, Via Lincoln 5 - 81100 Caserta, Italy}
\address[add25]{ELTE Eötvös Loránd University, Institute of Physics, Pázmány Péter sétány 1/A, Budapest 1117, Hungary
}

\date{Received: date / Revised version: date}

\begin{abstract}
We introduce the LUNA neutron detector array developed for the investigation of the $^{13}$C($\alpha$,$n$)$^{16}$O reaction towards its astrophysical $s$-process Gamow peak in the low-background environment of the Laboratori Nazionali del Gran Sasso (LNGS). Eighteen $^{3}$He counters are arranged in two different configurations (in a vertical and a horizontal orientation) to optimize neutron detection efficiency, target handling and target cooling over the investigated energy range E$_{\alpha,\text{lab}}=300-400$ keV (E$_{\text{n}}=2.2-2.6$ MeV in emitted neutron energy). As a result of the deep underground location, the passive shielding of the setup and active background suppression using pulse shape discrimination, we reached a total background rate of $1.23\pm0.12$ counts/hour. This resulted in an improvement of two orders of magnitude over the state of the art allowing a direct measurement of the $^{13}$C($\alpha$,$n$)$^{16}$O
cross-section down to E$_{\alpha,\text{lab}}=300$ keV. The absolute neutron detection efficiency of the setup was determined using the $^{51}$V(p,n)$^{51}$Cr reaction and an AmBe radioactive source, and completed with a Geant4 simulation. We determined a (34$\pm$3) \% and (38$\pm$3) \% detection efficiency for the vertical and horizontal configurations, respectively, for E$_{\text{n}}=2.4$ MeV neutrons.
\end{abstract}


\end{frontmatter}



\section{Introduction}
\label{S:1}

The $^{13}$C($\alpha$,$n$)$^{16}$O reaction is the dominant neutron source for the synthesis of the elements heavier than iron via slow neutron captures (the $s$-process) in thermally pulsing, low-mass AGB stars \cite{Gallino1998}. 
The relevance of this reaction for the synthesis of heavy elements and the most recent experimental studies are extensively illustrated in \cite{Drotleff1993, Heil2008, Cognata2012, Mukhamedzhanov2017, Cristallo2018}. As underlined by \cite{Cristallo2018}, direct data in the low energy region are highly desirable to better constrain the $^{13}$C($\alpha$,$n$)$^{16}$O astrophysical reaction rate.
In order to measure the rapidly declining cross-sections in this energy region, a low neutron background and high neutron detection efficiency are needed.

Here, we introduce the LUNA neutron detector array, which
has been developed for $^{13}$C($\alpha$,$n$)$^{16}$O reaction cross-section measurements at the underground LUNA 400kV accelerator \cite{Formicola2003} of the Laboratory for Underground Nuclear Astrophysics (LUNA) facility installed in the Laboratori Nazionali del Gran Sasso (LNGS), Italy.

The deep underground environment of the LNGS leads to a reduction of the natural neutron background  by three orders of magnitude with respect to the surface  \cite{Best2016, Arneodo1999}. At this level, the intrinsic radioactivity of the detector and other nearby materials becomes the dominant source of background \cite{HashemiNezhad:1998}. To constrain the astrophysical reaction rate, the measurement of the $^{13}$C($\alpha$,$n$)$^{16}$O reaction cross-section (Q value\,=\,2.216 MeV) needs to be performed at energies E$_{\alpha,\text{lab}} < 400$ keV \cite{Cristallo2018}.
At these beam energies, emitted neutrons are in energy range E$_{\text{n}}=2.2-2.6$ MeV, given in laboratory coordinate system, considering also the counters position of the LUNA neutron array.  
Assuming a beam current of I$_{\text{beam}}=100$ $\mu$A, the estimated neutron emission rate at these energies is as low as 1 neutron/hour. 
Therefore, our goal was to minimize the background counting rate and to optimize the absolute neutron detection efficiency ($\eta_{\text{n}}$) of the LUNA neutron detector array in the neutron energy region of around 2.4 MeV.

In contrast to $\gamma$-ray spectroscopy, the determination of the neutron efficiency curve is challenging mainly due to the limited choices of sources with accurately known energy spectra and/or angular distributions (in the case of reactions) and in some cases the limited availability of accurately calibrated sources.
A standard procedure is to employ radioactive sources ($^{252}$Cf, AmBe), which emit neutrons with a continuous energy spectrum, in combination
with Monte Carlo simulations \cite{Drotleff1993,Harissopulos2005,Drotleff1995}. The efficiency curve can be complemented using nuclear reactions, e.g. the $^{51}$V($\text{p,n}$)$^{51}$Cr reaction \cite{Deconninck1969, Falahat2013, Pereira2010, Wrean2000}.
To constrain the uncertainty of efficiency determination, the design of neutron detection setup should be optimized to obtain an energy-efficiency relation as flat as possible along the energy range of interest \cite{Pereira2010}. 

This paper consists of the following sections: section \ref{sec:setup} is devoted to the technical description of the neutron detector array; in section \ref{sec:background}, the background characterization of the experimental setup is presented; section \ref{exp_eff} describes the determination of the neutron detection efficiency, followed by an overview of the simulation in section \ref{sect:Geant_describe}; results are discussed in section \ref{section:result_discuss}.

\section{Description of the neutron detector array} \label{sec:setup}

The detector array contains eighteen $^{3}$He filled proportional counters with stainless steel housing\footnote{Manufactured by GE Reuter-Stokes, Inc., model numbers RS-P4-0816-217 and RS-P4-0810-250. The nominal filling pressure is 10 atm.}. They are arranged in two concentric rings around the target chamber: twelve counters of 40 cm active length are located at a radius of 11 cm, and six counters of 25 cm active length are located at 6 cm radius. This configuration allows for a nearly 4$\pi$ solid angle coverage around the target. 
As $^{3}$He has a very high cross-section for capturing thermal neutrons through the $^{3}$He(n,p)$^{3}$H reaction ($\sigma_{Thermal}=5330$ barn, $Q=764$ keV), effective thermalisation of the emitted neutrons from the $^{13}$C($\alpha$,$n$)$^{16}$O reaction is required. To achieve this, the counters are embedded in a high-density polyethylene (PE) moderator. 

\begin{figure*}[t]
\begin{center}
    \begin{tabular}{c c}

    \includegraphics[width=1\columnwidth]{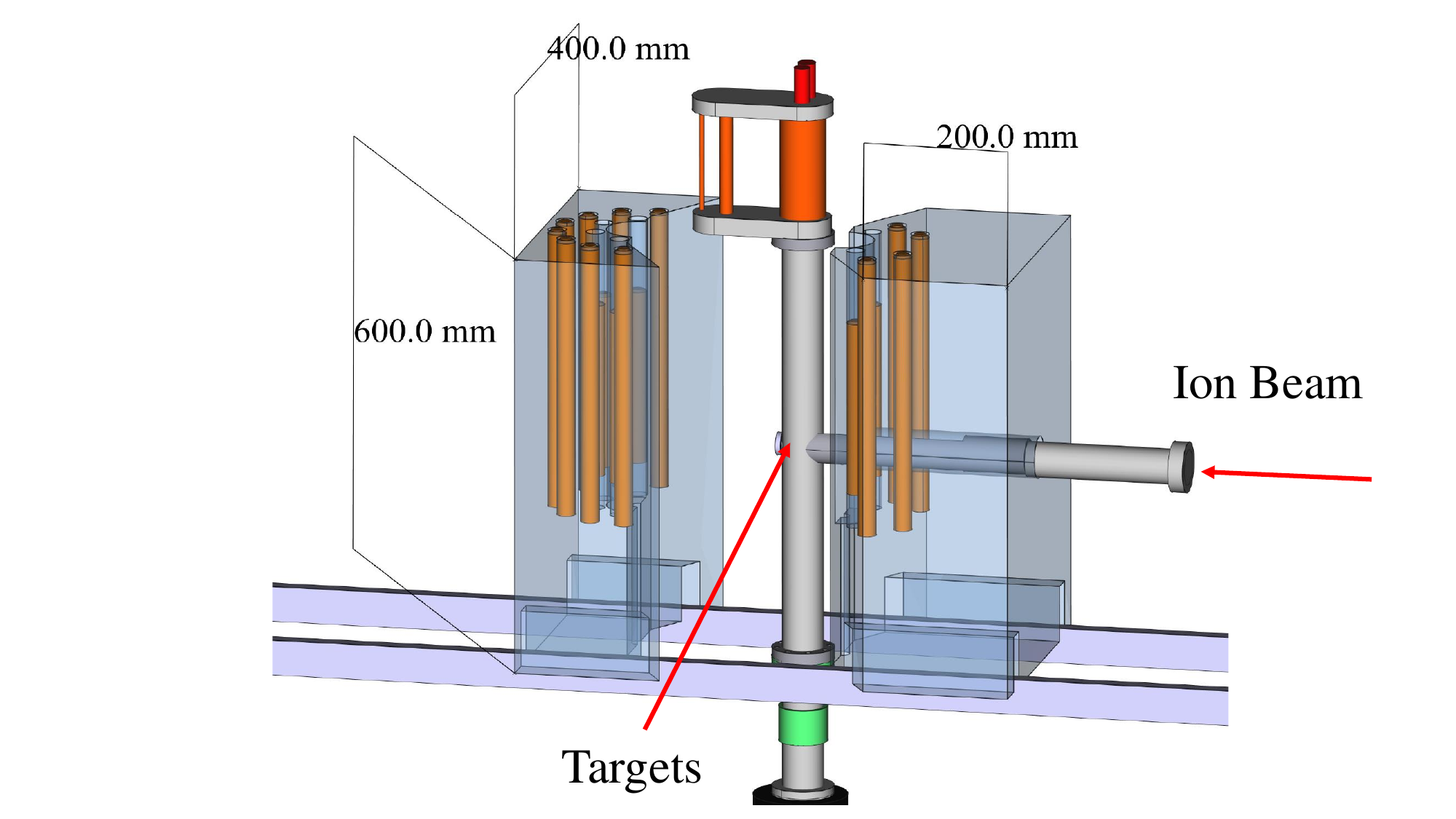}
    &
    \includegraphics[width=1\columnwidth]{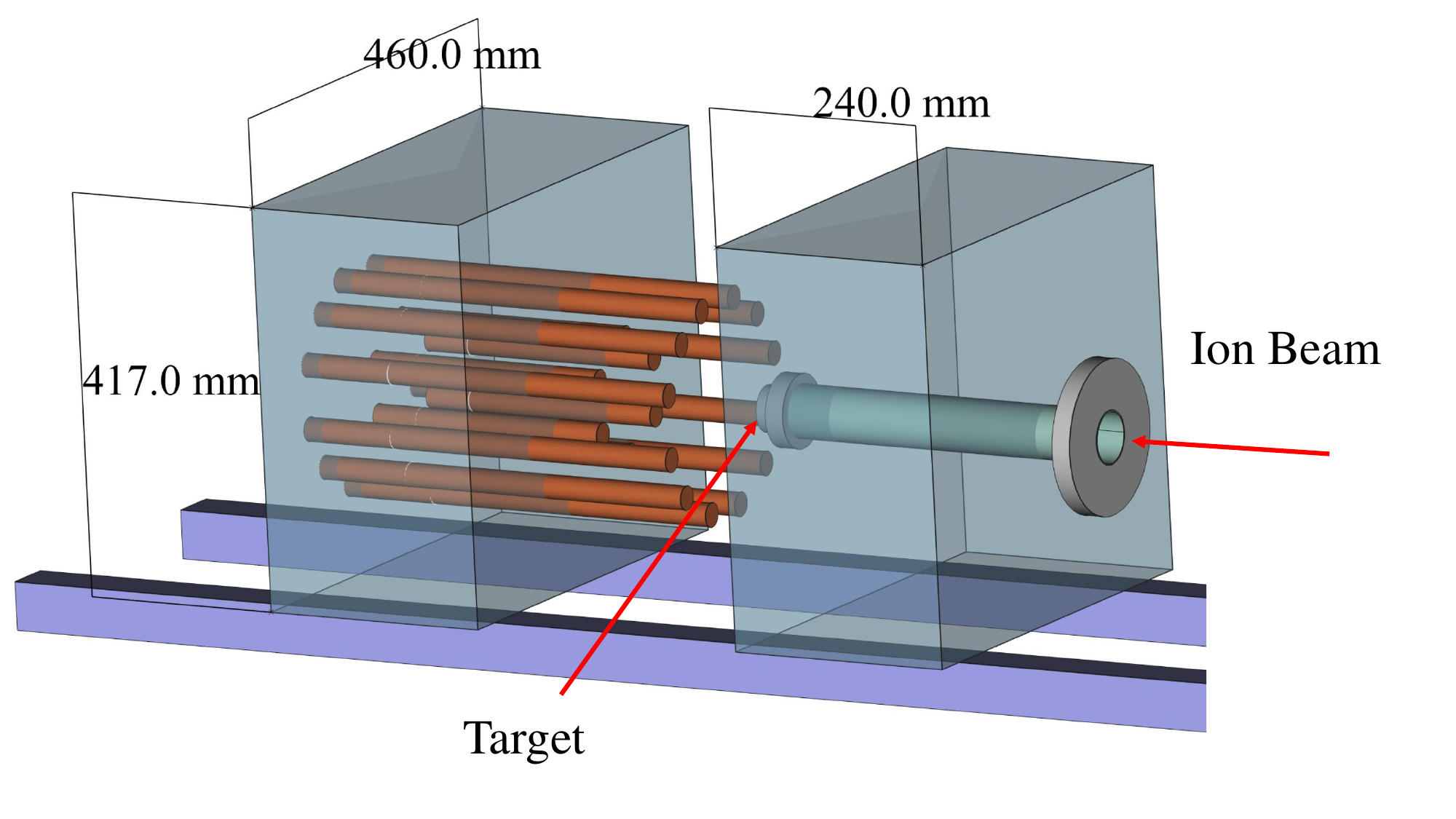}
    \end{tabular}
\end{center}
\caption{Vertical (left panel) and horizontal (right panel) setup of the LUNA neutron detector array. Dimensions of the moderators, target positions and beam direction are also indicated (see text for more details).}
\label{fig:Design_of_setup_MT}
\end{figure*}

To measure the very low number of neutrons emitted at the lowest energies, the neutron detection efficiency, target handling, and active target cooling has to be optimized. Therefore, two detector geometries were designed for the experimental $^{13}$C($\alpha$,$n$)$^{16}$O campaign: in one configuration the counters positioned at 90 degrees to the beam axis and in the other the counters are positioned parallel to the beam (these two configurations are referred hereafter as $``$vertical$"$ and $``$horizontal$"$ setups, respectively, see figure~\ref{fig:Design_of_setup_MT}) were used. Moreover, the distribution of the $^{3}$He counters was also optimized to maximize the efficiency at the energy of interest (around E$_{\text{n}}=2.4$ MeV).

The vertical setup was used with a $``$T$"$-shaped stainless steel target chamber composed of vertical and horizontal tubes with an outer diameter of 52 mm. The beam arrives along the horizontal tube of the chamber and impinges on the target mounted on a multi-stage, water cooled target holder with capacity for three targets installed in the vertical tube of the chamber. This minimizes possible contamination and simplifies the target changing procedure during the experiment. Moreover, the target holder could be extracted from the chamber without interfering with the moderator contributing to the stable condition of the setup. 

In the horizontal setup, a single, water cooled target was mounted at the end of the horizontal tube (a diameter of 40 mm). This setup allows for a higher neutron detection efficiency and better cooling capacity, and was used for the low-energy measurements where detection efficiency was of particular importance. The moderator is divided vertically in two, movable section to allow for easy exchange of the target (see figure~\ref{fig:Design_of_setup_MT}). 

In addition, the moderator was surrounded with 25.4 mm (vertical configuration) and 50.8 mm (horizontal configuration) thick layer of 5 \% borated polyethylene (BPE) to further reduce the environmental neutron background. 
The shape of the polyethylene shielding in the horizontal and the vertical setup was constrained by the arrangement of the $^{3}$He tubes, which was optimized taking into account the target changing procedure, efficiency and target cooling capacity.
The neutron counters in the vertical setup was used to measure the $^{13}$C($\alpha$,$n$)$^{16}$O reaction cross-section in the energy range E$_{\alpha,\text{lab}}=360-400$ keV, while higher neutron detection efficiency, further background suppression by more shielding material, and better cooling of the horizontal setup allowed for the low-energy measurements at E$_{\alpha,\text{lab}}=300-360$ keV.

In both setups, a cooling loop running deionised water at 5\,$^\circ$C is integrated in the target holders for beam power dissipation (on the order of $\sim$\,100 W).
Further descriptions of the setup can be found in previous publication \cite{CianiPhD}.

\subsection{Data acquisition system} \label{DAQ}

Signals from the counters were shaped in CAEN A1422 charge sensitive preamplifiers  with a gain of 90 mV/MeV. Each module has 8 channels plus a common test input, for a total of 3 modules used. The signal is fed to a CAEN V1724 100 MS/s 14-bit digitizers, which are read out through a common USB connection. The 18 signals (one from each $^{3}$He counter) are distributed to three digitizers occupying 6 channels/module. 

A pulse height analysis firmware implementing a trapezoidal shaping algorithm (CAEN Dpp-PHA) \cite{Jordanov:1994} was used for the efficiency measurements (see Section \ref{exp_eff_site_atomki} and \ref{exp_eff_Naples}). To estimate the system dead time, a reference pulse generated with a BNC DB-2 random pulse generator was fed into the test input of each preamplifier and one of the free channels in each digitizer. 

For the underground measurements, the signals of each counter were read out and digitized for off\,-\,line analysis using a custom Labview interface \cite{CAEN-libs}. This configuration allows to perform pulse-shape discrimination \cite{Balibrea2019}, which is of particular importance for the low-energy measurements at $E_{\alpha,\text{lab}}<340$ keV (see Section \ref{sec:background}). Due to the low event rates ($<$1 Hz), the underground measurements were considered dead-time free.

\section{Internal and external backgrounds} \label{sec:background}

\begin{figure}[htb!]
    \centering
    \includegraphics[width=\columnwidth]{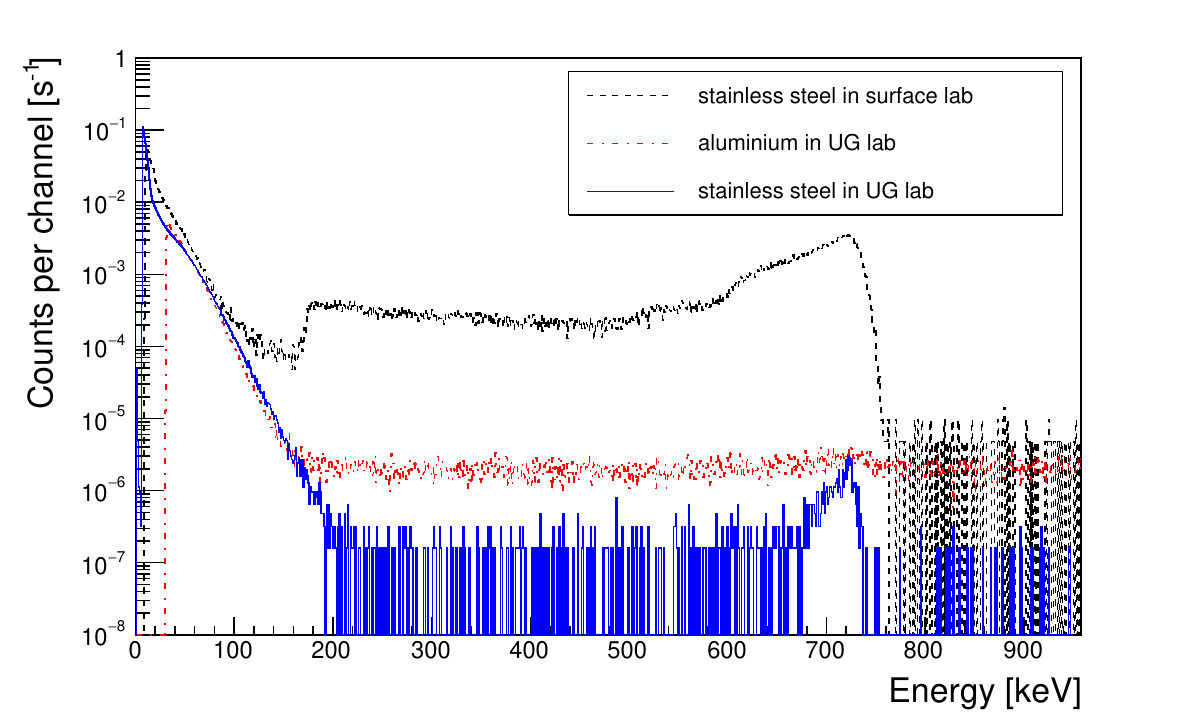}
    \caption{Comparison of background spectra of bare $^{3}$He counters acquired on the surface and in the underground laboratory (UG lab) using single counters made from aluminum (dash-dotted line) and from stainless steel (dashed and solid lines). }
    \label{fig:neutron_spectra}
\end{figure}

As already pointed out, both high detection efficiency and low background are crucial to achieve the required sensitivity for a low energy measurement of the $^{13}$C($\alpha$,$n$)$^{16}$O reaction. In our case, the location of the experimental apparatus and the properly selected material of the enclosure of the $^{3}$He counters imply an unique low-natural background.
The comparison of the experimental background in the signal region of interest is presented in figure~\ref{fig:neutron_spectra}: showing data taken on the Earth's surface (dashed line), in the underground laboratory of the LNGS (dash-dotted line) using single counter made of aluminium, and in the underground laboratory using counter made from stainless steel -- part of the setup we are describing here (solid line). 

The signals from the $^{3}$He(n,p)$^{3}$H products (proton and triton) appear between 150 and 800 keV including the full-energy deposition (764 keV) and the cases when one of the product is absorbed in the wall of the counters, so called wall-effect. All three measurements were performed without any moderator or shielding around the detector.

The alpha activity of radioactive contaminants (from the decay chain of uranium and thorium) inside the enclosure of the counters leads to an intrinsic background. From the count rate in the energy region above the thermal neutron peak and assuming a flat alpha spectrum \cite{Best2016}, the alpha activity\footnote{Calculated as the total alpha yield, assuming a flat alpha spectrum also in the neutron signal region, measured for the entire setup divided with the total surface area (in cm$^{2}$) of the $^{3}$He counters.} of the enclosure material was evaluated to be ($2.71\pm0.07) \cdot 10^{-6}$ $\alpha$ $\text{cm}^{-2}\text{s}^{-1}$ (to be compared to $6 \cdot 10^{-5}$ $\alpha$ $\text{cm}^{-2}\text{s}^{-1}$ for the Al counters \cite{Best2016}). The average alpha background rate integrated over the neutron signal region was about $1.50\pm0.04$ $\alpha$ $\text{hour}^{-1}$ for the entire array.

For the entire setup, the total background rate inside the neutron signal region (between 200 keV and 800 keV in figure \ref{fig:neutron_spectra}) is $3.34\pm0.11$ (vertical) and $3.08\pm0.09$ (horizontal) counts/hour. This is the sum of the intrinsic alpha and the extrinsic neutron backgrounds. 

However, going towards the $s$-process Gamow peak, the reaction yield of the $^{13}$C($\alpha$,$n$)$^{16}$O reaction drops to the 1 event/hour level. 
Therefore, even the achieved very low background severely limits the sensitivity towards the lowest energies and needs to be further suppressed. For this, we applied the pulse-shape discrimination method described in ref. \cite{Balibrea2019} to the signals from the detectors, removing practically all $\alpha$-particle signals and lowering the total background to $1.23\pm0.12$ counts/hour. It is worth noting that this number is affected by the efficiency of PSD method as it is described in ref. \cite{Balibrea2019}, which we will consider in the cross-section determination. The achieved background rate represents an improvement of two orders of magnitude over similar setups \cite{Drotleff1993,Harissopulos2005}.

\section{Efficiency measurement} \label{exp_eff}

To determine the neutron detection efficiency ($\eta_{n}$) of both configurations of the LUNA neutron detector array, we performed measurements using the $^{51}$V($\text{p,n}$)$^{51}$Cr reaction at the Van de Graaff Laboratory of Institute for Nuclear Research (Atomki) in Debrecen, Hungary. To further extend the investigated neutron energy region, we also performed an efficiency campaign using an AmBe neutron source (described in Section \ref{exp_eff_Naples}) at the University of Naples ``Federico II''. The experimental arrangement of the neutron arrays in both efficiency campaigns was identical with one of the underground measurements, unless indicated otherwise in the respective section.

\subsection{The $^{51}$V($p$,$n$)$^{51}$Cr reaction}\label{exp_eff_site_atomki}

The measurements were performed at the 30$^\circ$ beamline of the 5\,MV Van de Graaff accelerator of Atomki\cite{VdGAtomki}. 
The beam was transported through a series of tantalum collimators resulting in a $\approx$\,5\,mm diameter beamspot on the target surface. The closest collimator was located 50\,cm from the target outside the volume of the moderator. 
The accumulated beam charge was measured by an ORTEC 439 Digital Current Integrator.

The $^{51}$V($\text{p,n}$)$^{51}$Cr reaction (Q\,=\,-1534.8\,keV) was used to produce monoenergetic neutrons in the energy range below 1\,MeV. Due to the slow variation of neutron intensity and energy with angle, combined with the well-known target preparation and utilization, the $^{51}$V(p,n)$^{51}$Cr reaction is widely used for calibration of neutron detectors \cite{Deconninck1969, Falahat2013, Pereira2010, Ramstrom1976, Lund1980}.
However, its application is limited by the opening of additional neutron channels above E$_\text{p,lab}=2330$ keV, which corresponds to the first excited state of $^{51}$Cr at E$_{x}=749$ keV. Above this energy, neutrons of different energies (n$_{0}$ for ground state transition, n$_{1}$ for the first excited state, etc.) might be mixed in the emitted neutron spectrum of the reaction. In spite of the n$_{1}$ neutron channel opening at E$_\text{p,lab}=2330$ keV, the work of \cite{Deconninck1969} indicates a negligible contribution of this neutron group to the total flux up to E$_\text{p,lab}=2600$ keV.
Therefore, we performed the measurements at E$_\text{p,lab}=1700, 2000, 2300, 2600$ keV assuming monoenergetic n$_{0}$ neutrons emitted at 130, 420, 710 and 990\,keV energy, respectively. 

The targets for these measurements were made by evaporating natural vanadium with nominal thicknesses of 60\,-\,200\,$\mu$g/cm$^{2}$ (typically $7-17$ keV proton energy loss at E$_{\text{p,lab}}=1700-2600$ keV) onto 0.3 mm thick tantalum backings. The beam intensity was limited to minimize the dead time of the data acquisition system. The average beam intensity varied between 20 nA and 500 nA. A blank tantalum target was also irradiated before each vanadium target run in order to investigate the possible contribution of beam induced background to the measured neutron yield. It was always found to be less than 0.1 \%.

The determination of the total number of emitted neutrons was based on the activation technique \cite{gyu19}. The $^{51}$V($\text{p,n}$)$^{51}$Cr reaction emits neutrons and produces an equivalent number of $^{51}$Cr radioactive nuclei. These nuclei decay via electron capture with a half-life of t$_{1/2}=27.7025(24)$ days. With a branching ratio $B$ of 9.91(1) \%, the decay leads to the first excited state in $^{51}$V and is followed by the emission of a 320\,keV $\gamma$ ray \cite{Yalcin2005}. The detection of this $\gamma$ ray provides the possibility of determining the number of reactions and hence the number of neutrons produced. 

After irradiation, off-line $\gamma$-ray measurements of the activated targets were performed at Atomki using a 100 \% relative efficiency HPGe detector placed in a complete 4$\pi$ lead shielding \cite{Halasz2012}. The absolute efficiency ($\eta_{320}$) of the HPGe detector was (13.4$\pm$0.4) \% at E$_{\gamma}$=320\,keV, as determined with calibrated radioactive sources of $^{137}$Cs, $^{60}$Co, $^{133}$Ba and $^{152}$Eu. The efficiency curve of this particular detector and the method of the efficiency calibration is presented in \cite{gyu19} Section 3.1. 

The neutron detection efficiency, $\eta_{\text{n}}$, of the detector array at a given incident proton energy\footnote{The energy of the emitted neutrons as a function of emission angle relative to the incident beam direction is calculated based on the equation C.5. in ref. \cite{Iliadis} and the arithmetic mean is used as an average neutron energy assigned to the incident proton energy.} can be calculated using the standard formula of activation \cite{gyu19}. The number of reactions (N$_{\text{R}}$) that take place during the activation time $t_i$ can be obtained as:
\begin{equation}
   N_{R}=\frac{N_{\gamma}}{{B\cdot\eta_{320}}}\cdot\frac{e^{\lambda t_w}}{1-e^{-\lambda t_c}}\cdot\frac{\lambda\cdot t_i}{1-e^{-\lambda t_i}}, 
\label{eq:51V}
\end{equation}
where N$_{\gamma}$ is the number of detected 320\,keV $\gamma$ rays (after dead time correction), $t_c$ is the counting time, $t_w$ is the waiting time elapsed between the end of the irradiation and the start of the counting, and $\lambda=\text{ln}(2)/\text{t}_{1/2}$ is the decay constant of $^{51}$Cr.

Then $\eta_{n}$ can simply be calculated as:
\begin{equation}
    \eta_{n}=\frac{N_{n}}{N_{R}},
\end{equation}
where N$_{n}$ is the number of the detected neutron events. For the determination of N$_{n}$, dead time correction is applied using the signal of a pulser with a rate of 20 pulses/sec.

\subsection{Measurement with an AmBe source}  \label{exp_eff_Naples}

To extend the efficiency towards higher neutron energies, another campaign of measurements was performed using an AmBe neutron source (manufactured by Eckert \& Ziegler). As only a nominal activity was known, we performed the calibration campaign as follows.

The combination of $^9$Be and the $\alpha$-emitter $^{241}$Am leads to the production of neutrons with energies up to 12 MeV via the reaction 
$$
^{9}\text{Be}+^{4}\text{He} \rightarrow ^{12}\text{C}+n \;,
$$
leaving $^{12}$C in either the ground or the excited state. For n$_{1}$ emission, a $\gamma$ ray of E$_{\gamma}$=4.4 MeV is emitted from the de-excitation of the first excited state in $^{12}$C. The probability of n$_{1}$ emission, or the $\gamma(4.4 $MeV$)$/n$_{total}$ ratio is $\text{R}=0.575\pm0.028$ (see \cite{Liu2007} and references therein). Therefore, by measuring the $\gamma$-activity we can determine the total activity of the source.

To better constrain the systematic uncertainty of the $\gamma$-ray measurements, three different types of detectors (LaBr$_3$:Ce, NaI, HPGe) were used simultaneously. The detectors were arranged at $\theta=30^\circ, 330^\circ$ and $180^\circ$ with respect to the surface of the neutron source. The source was placed on a target holder in the center, at a distance of 10 cm from the detectors. To suppress the low-energy gammas emitted by the source, lead plates with 1 mm thickness were inserted between the source and the $\gamma$-detectors.
The absolute efficiency ($\eta_{\gamma}$) curve of the different detectors was determined using $^{60}$Co and $^{56}$Co radioactive sources (activities $A_{\gamma}$ are known at a precision of $\leq$1 \% and 3 \%, respectively) and extended to higher energies using an extrapolation that was cross-checked with a Geant4 simulation. Overall a relative uncertainty of the extrapolated efficiency of $\sim 3 \%$ was reached at $E_{\gamma}=4.4$ MeV.

\begin{figure}[htb!]
    \centering
    \includegraphics[width=\columnwidth]{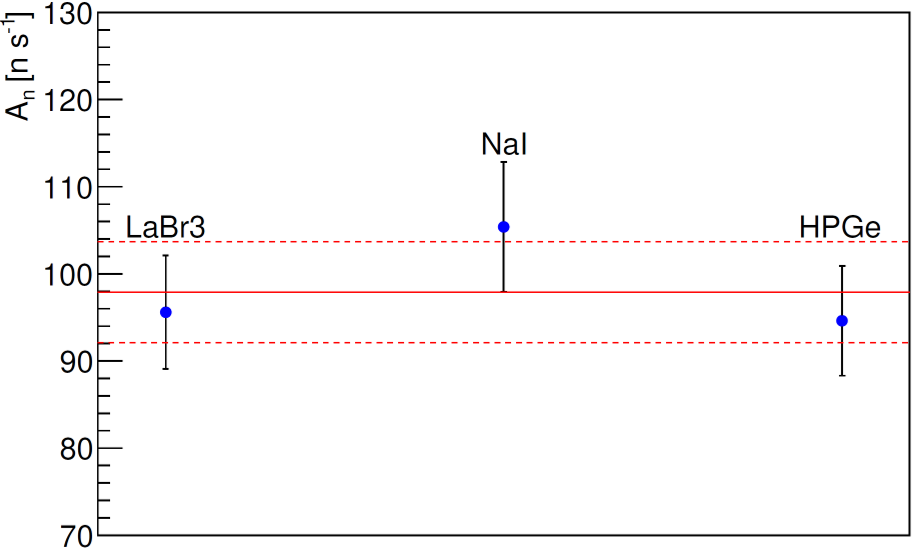}
    \caption{Comparison of the neutron activities evaluated with the three different $\gamma$-ray detectors. The solid and the two dashed lines represent the weighted average and the associated one sigma uncertainty range.}
    \label{fig:neutron_spectra2}
\end{figure}

The activity A$_n$ of the source can be determined through
\begin{equation}
    A_n=\frac{N_\gamma}{\eta_{4.4}Rt},
\end{equation}
where $N_{\gamma}$ is the number of net counts in the full-energy peak. The measurement time is defined by $t$ and $\eta_{4.4}$ is the photo-peak efficiency of the $\gamma$-ray detectors at E$_{\gamma}=4.4$ MeV. The A$_{\text{n}}$ values and their weighted mean activity determined through the different $\gamma$-ray detectors are presented in figure~\ref{fig:neutron_spectra2}.
The activities measured with the three different $\gamma$-ray detection setups agree within one sigma. 
We obtained a A$_{\text{n}}=97.9\pm5.8$ n/s as a weighted average of the values. The indicated value takes also into account the uncertainty of the branching ratio R (4.8 \%). 

Since the AmBe neutron spectrum has a wide energy range (E$_{\text{n}}$\,=\,0 to 12\,MeV), the  effective (or weighted average) neutron energy E$_{\text{n-AmBe}}$ was determined using the neutron emission probability at a given energy E$_{\text{n}}$ and the energy dependence of the efficiency.  
This efficiency trend was calculated using Geant4 simulation (see in Section \ref{sect:Geant_describe}) and the reference neutron spectrum was taken from ISO 8529-2 \cite{AmBe_1,AmBe_2}. The resulting value was E$_{\text{n-AmBe}}=4.0\pm0.3$ MeV.

The physical quantities, parameters and their uncertainties in the experimental efficiency calculation using the $^{51}$V(p,n)$^{51}$Cr reaction and the radioactive source are reported in table~\ref{tab:Uncertainty_MT}.
The total uncertainties are obtained from the quadratic sum of partial uncertainties unless indicated otherwise in the respective section.

\begin{table*}[t]
\caption{Physical quantities, parameters and their uncertainties applied in the experimental neutron detection efficiency calculations.}
\begin{center}
\resizebox{1\columnwidth}{!}{ \begin{tabular}{p{2.6cm}| p{1.4cm} | p{1.0cm} p{2.3cm} }\hline
    Neutron emitters & Parameters & Value & Uncertainty (\%) \\ \hline
      $^{51}$V(p,n)$^{51}$Cr & T$_{1/2}$ (day) & 27.703 & 0.01  \\
      & B & 0.0991 & 1  \\
      & $\eta_{320}$ & 0.134 & 3   \\
      & N$_{\gamma-AA}$ & - & $\leq$2  \\ \hline
      AmBe & $\eta_{4.4}$ & - & 3 \\
      & $A_{\gamma}$ ($^{60}$Co) & - & $\leq$1\\& $A_{\gamma}$ ($^{56}$Co) & - & 3 \\
      & R & 0.576 & 4.8 \\ \hline
      \end{tabular}}
\end{center}
\label{tab:Uncertainty_MT}
\end{table*}

\section{Simulation of the detector response} \label{sect:Geant_describe}

We used the Geant4 toolkit \cite{Allison2016,Agostinelli2003} \footnote{Geant4 version 10.03, with $``$neutron high precision$"$ physics and thermal scattering corrections enabled for water and polyethylene.} to simulate the detector response. All materials in the detector region were included according to the technical drawings of the various components (moderator, shielding, detectors, target chamber, water cooling). The simulation was used to calculate the relative distribution of counting rates between the counters, the ratio between total yields of the outer and inner ring, and the energy dependence of the neutron detection efficiency.
The energy broadening and angular distribution effects due to the transformation from the center-of-mass frame to the laboratory frame were taken into account for the nuclear reaction measurements. The neutron emission of the AmBe source was assumed to be isotropic.

\begin{figure}[htb!]
\begin{center}
    \begin{tabular}{c c}
    \includegraphics[width=1\columnwidth]{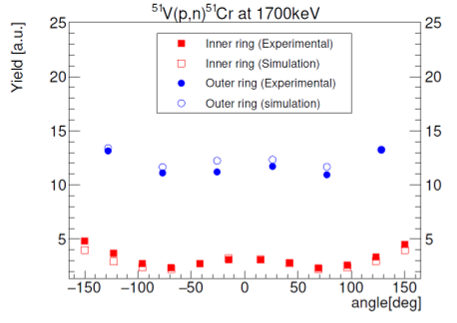}\\
    
    \includegraphics[width=1\columnwidth]{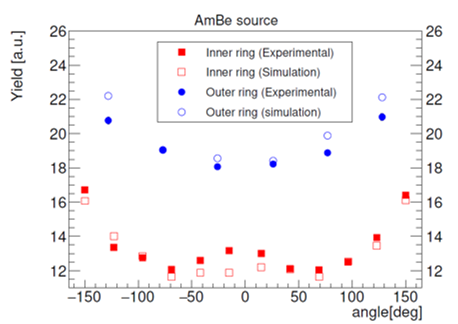}\\
    \end{tabular}
\end{center}
\caption{Comparison between the experimental (filled symbols) and the simulated (empty symbols) yields as a function of counters position. Circles (squares) refer to the inner (outer) ring of $^{3}$He counters in vertical configuration. Experimental data are obtained by exploiting the reaction $^{51}$V(p,n)$^{51}$Cr at E$_{\text{p,lab}}$\,=\,1700 keV (top panel) and by using the AmBe source (bottom panel). 
}
\label{fig:Yield_distribution}
\end{figure}

Simulations of the hit patterns in the vertical configuration were carried out for the $^{51}$V$(\text{p,n})^{51}$Cr reaction at E$_{\text{p,lab}}=1700, 2000, 2300$ and 2600\, keV, corresponding to average neutron energies of E$_{\text{n}}=130, 420, 710$ and 990\,keV; and for the AmBe neutron source, corresponding to E$_{\text{n}}=4000\pm300$ keV, respectively. 
The comparison between measurements and simulation at E$_{\text{p,lab}}=1700$ keV and using the AmBe source is shown in figure~\ref{fig:Yield_distribution}, with the beam axis at 0 degree. 

It is worth noting that the total number of detected neutrons from the simulation was normalized to that obtained in the experimental measurements for the comparison. Due to the small statistical uncertainties, the error bars are not visible in the figure.

There is a fair agreement between simulation and experimental measurements. The not-uniform distribution of the yields for the counters at different positions is also well reproduced by the simulation. This anisotropy is attributed to geometrical effect, the distribution of construction materials (e.g., the water in the cooling loop), and the kinematics of the nuclear reaction. 

Therefore, we consider our simulation model suitable to obtain $\eta_{\text{n}}$ in the LUNA energy range.

\section{Results and discussion} \label{section:result_discuss}

To determine the neutron detection efficiency $\eta_{\text{n}}$ at the neutron energy region of interest for the $^{13}$C($\alpha$,$n$)$^{16}$O measurement (E$_{n}\sim$2.4 MeV), the experimental efficiency data were compared with the simulated results (including the kinematic energy distribution effect, see section \ref{sect:Geant_describe}).
The measured and simulated $\eta_{\text{n}}$ obtained from the $^{51}$V(\text{p,n})$^{51}$Cr reaction and AmBe neutron source, and the energy of bombarding particle and emitted neutrons in keV are summarized in table~\ref{tab:EFF_data}.
A notable discrepancy is present between the experimental and simulated data. Therefore, we determined a scaling factor L$_{\text{scal}}=0.77\pm0.01$ between the simulated and the measured efficiency using the least squares method. The application of such a scaling factor is commonly used to compensate the deficiencies of the simulation model \cite{Li2018, Ende2016, Lemrani2006} and the absolute scale is consistent with those found with similar neutron detection setups \cite{Falahat2013, Pereira2010}. The rescaled simulated data are also presented in table~\ref{tab:EFF_data}.

\begin{table*}[t]
\caption{Experimental and simulated neutron efficiencies for the different detector geometries given in \%. The presented simulated results are already corrected for reaction kinematics and angular distribution. $``$Simulated-rescaled$"$ represents the calculated efficiency ($``$Simulated$"$) by the Geant4 code is scaled to the $``$Experimental$"$ data using the least squares method.} 
\begin{center}
\begin{tabular}{c | c c | c c | c c | c c}\hline
   Reaction & E$_{\text{p}}$(keV) & E$_{\text{n}}$(keV)* & \multicolumn{2}{c|}{Experimental} & \multicolumn{2}{c|}{Simulated} & \multicolumn{2}{c}{Simulated-rescaled}   \\
    &  &  & vert. &  hor. & vert. & hor. & vert. & hor. \\ 
   \hline
$^{51}$V(\text{p,n})$^{51}$Cr	&	1700	&	130$\pm$20	&	42.3$\pm$	1.4	&	46.6$\pm$	2.3	&	56.4	&	63.0	&	43.7	&	48.1	\\	
	&	2000	&	420$\pm$30	&	41.2$\pm$	1.3	&	45.4$\pm$	2.3	&	54.2	&	61.0	&	42.0	&	46.6	\\	
	&	2300	&	710$\pm$50	&	40.4$\pm$	1.3	&	44.0$\pm$	2.2	&	52.6	&	58.2	&	40.8	&	44.5	\\	
	&	2600	&	990$\pm$60	&	38.5$\pm$	1.2	&			&	50.3	&		&	39.0	&		\\	\hline
AmBe	&		&	4000$\pm$300**	&	32.4$\pm$	1.6	&	35.5$\pm$	1.7	&	35.8	&	40.6	&	27.7	&	31.2	\\

      \end{tabular}
\end{center}
\footnotesize $*$ Due to the reaction kinematics and finite target thickness, neutrons are not strictly monoenergetic. Here, the average energy and the minimum and maximum energies are given.\\
$**$ The AmBe source emits neutrons with energies between E$_{\text{n}}=0-12$ MeV. The indicated value is the effective neutron energy (see Section \ref{exp_eff_Naples}).
\label{tab:EFF_data}
\end{table*}

\begin{figure}[hbt!]
\begin{center}
\begin{tabular}{c c}
\includegraphics[width=0.96\columnwidth]{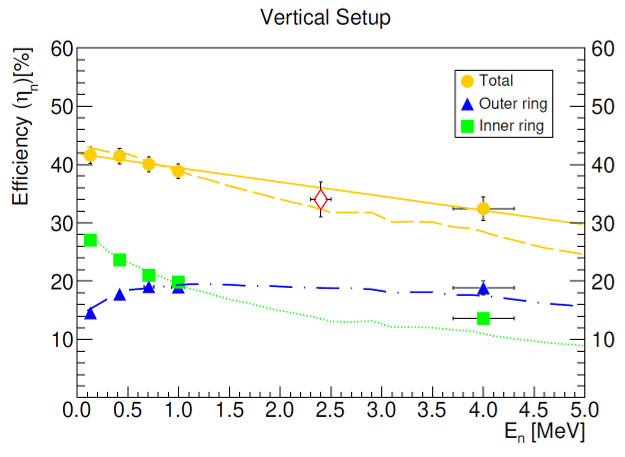}\\
\includegraphics[width=0.96\columnwidth]{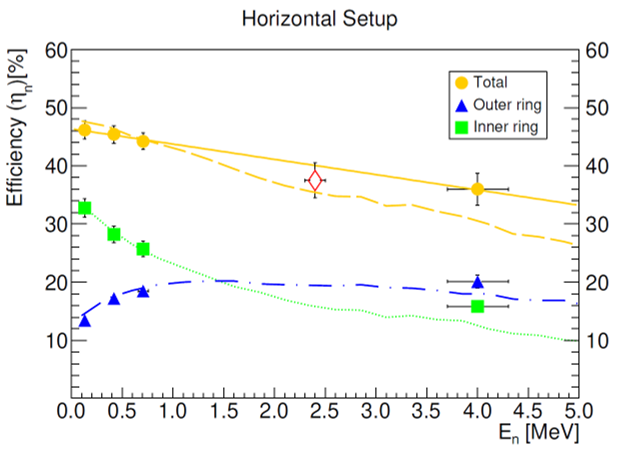}\\
\end{tabular}
\end{center}
\caption{Experimental efficiencies (filled symbols) and the rescaled-nominal efficiency curve (dashed line) obtained using the vertical (upper panel) and horizontal (lower panel) setups, respectively. The simulated and the experimental efficiencies related to the inner (squares and dotted lines) and outer (triangles and dash-dotted lines) ring of the setups are also presented. The linear fit of the experimental data (solid lines) and the proposed efficiency value at E$_{\text{n}}=2.4$\,MeV (empty diamonds) are also presented. 
}
\label{fig:comparison}
\end{figure}

L$_{\text{scal}}$ can be used to calculate the nominal efficiency curve\footnote{This is the efficiency curve assuming monoenergetic, isotropically distributed neutron emission in the laboratory frame.}. 
This is presented in figure~\ref{fig:comparison} (dashed line) together with the experimental results (filled symbols), where the plotted experimental data were corrected with the kinematic energy distribution effect to obtain the nominal efficiency values. To cross-check the consistencies of our model, the results for the inner (dotted line) and outer (dash-dotted line) rings of the detector array are shown separately. 

A general agreement was obtained between the simulated and the experimental data. However, there is a slightly different energy dependence of the efficiency curve between the experimental and simulated data set, which may be attributed to various sources of uncertainty in the simulation Monte Carlo code, such as geometry, multiple-scattering of neutrons on the construction materials, elastic, inelastic
scattering and nuclear reaction cross sections, molecular vibrational and rotational
excitation modes in the moderator material.

Because the relevant energy range
of this measurement is relatively narrow and well defined compared to the energy
range where the efficiency is measured and simulated, with the quoted uncertainties
the determined efficiency seems robust, regardless the possibly energy dependent
scaling factor (see below). Moreover, in the next future the installation of LUNA MV facility at LNGS will allow to extend the efficiency measurements adding more high energy calibration points by other reaction or by a calibrated $^{252}$Cf source.
This will permit to better constrain energy dependence.

The simulation results in efficiencies of (35$\pm$1) \% (horizontal) and (32$\pm$1) \% (vertical) for the two configurations of the neutron detector array in the LUNA energy range (E$_{\alpha,\text{lab}}=300-400$ keV, E$_{\text{n}}\sim$2.4\,MeV).

To obtain a more robust $\eta_{\text{n}}$ in the LUNA energy range, we also used another approach to interpolate the $\eta_{\text{n}}$ from the low\,- and high-energy data points.
The experimental efficiencies as a function of neutron energy (corrected with the kinematics energy distribution) were fitted with a linear function, as shown the solid lines in figure~\ref{fig:comparison}. The uncertainty of the interpolated data was estimated using random sampling of the measured data assuming Gaussian distribution of the assigned uncertainties and considering also the uncertainty of the AmBe effective neutron energy. The $\eta_{\text{n}}$ was interpolated to be (40$\pm$1) \% (horizontal) and (36$\pm$1) \% (vertical).
A relative $\approx$15 \% discrepancy is present between the interpolated and the simulated values. 
Therefore, we adopt the average of efficiency between the two methods with an assigned uncertainty to cover both the simulated and the interpolated value, (38$\pm$3) \% and (34$\pm$3) \% \footnote{The systematic uncertainties attributed to possible geometrical effect, such as the not point like\,-\, and off centered\,-\, beamspot, and asymmetrically located counters along the horizontal and vertical axis were controlled using Geant4 simulation and experimental measurement using AmBe source. A relative $\leq$1 \% was obtained on the $\eta_{\text{n}}$.} (indicated the red empty diamonds in figure~\ref{fig:comparison}) for the two setups in the LUNA energy range. 
It is worth noting that the values indicated above assume an isotropic distribution of the emitted neutrons. An energy-dependent correction should be applied to take into account the effect of the angular distribution of the neutrons emitted in the $^{13}$C($\alpha$,$n$)$^{16}$O reaction when calculating the reaction cross-section. This will be presented in forthcoming publication. 

\section{Summary}\label{section:summary}

A new LUNA neutron detector array has been commissioned for the investigation of the $^{13}$C($\alpha$,$n$)$^{16}$O reaction towards its astrophysical $s$-process Gamow peak in the low-background environment of the LUNA experiment at the Laboratori Nazionali del Gran Sasso. 

As a result of the low intrinsic activity of the $^{3}$He counters, the passive shielding of the experimental apparatus, the applied pulse shape discrimination technique and the deep underground location of the experiment, we achieved a total background rate $1.23\pm0.12$ counts/hour. 

Due to the large reduction of the reaction yield of the $^{13}$C($\alpha$,$n$)$^{16}$O reaction over the energy range $E_{\alpha,\text{lab}}=300-400$ keV investigated by the LUNA experiment, two different configurations of the setup were used with vertically and horizontally arranged $^3$He counters. This provided an optimized neutron detection efficiency, target handling and active target cooling.

We determined the neutron detection efficiency in a wide neutron energy range ($0.1-4.0$ MeV) experimentally using the $^{51}$V(p,n)$^{51}$Cr reaction and an AmBe neutron source combined with a robust simulation based on Geant4 code. A (38$\pm$3) \% (horizontal) and a (34$\pm$3) \% (vertical) absolute neutron detection efficiency of the setup were obtained as averages in the E$_{\text{n}}=2.2-2.6$ MeV range corresponding to the neutron emission of the $^{13}$C($\alpha$,$n$)$^{16}$O reaction in the LUNA experiment. 

We conclude that the total environmental neutron background of the LUNA neutron detector array in the deep underground location of the LNGS, combined with high neutron detection efficiency, creates a unique possibility to measure the cross-section of the $^{13}$C($\alpha$,$n$)$^{16}$O reaction approaching its $s$-process Gamow peak.
 
\section*{Acknowledgement}
 
The authors would like to thank the mechanical workshops of LNGS and INFN Naples for technical support. D. Ciccotti greatly helped during all aspects of the execution of this work.
Support from the National Research, Development and Innovation Office NKFIH (contract numbers PD129060, K120666 and NN128072), the University of Naples - Compagnia di San Paolo Program $``$STAR$"$, STFC-UK, DFG (BE 4100/4-1) and COST (ChETEC CA16117) is also acknowledged.
This work was also supported by the Polish National Science Centre under Contract No. 2014/14/M/ST2/00738 (COPIN-INFN Collaboration).

\bibliographystyle{model1-num-names}
\biboptions{sort&compress}
\bibliography{mainArxiv}

\end{document}